\documentstyle[preprint,floats,aps]{revtex}
\begin{document}
\draft
\title{Exact Solution of the Ising Model on Group Lattices \\
of Genus $g>1$}
\author{Tullio Regge and Riccardo Zecchina}
\address{Dipartimento di Fisica, Politecnico di Torino,\\
I-10129 Torino, Italy}
\maketitle

\begin{abstract}
We discuss how to apply the dimer method to Ising models on group lattices
having non trivial topological genus $g$. 
We find that the use of group extension and the existence of both
external and internal group isomorphisms greatly reduces the number of
distinct Pfaffians and leads to explicit topological formulas for their sign
and weight in the expansion of the partition function. The complete solution
for the Ising model on the Klein lattice group $L(2,7)$ with $g=3$ is given.
\end{abstract}


\section{Introduction}

Among the known approaches to the exact evaluation of the $2D$ Ising
partition function which have followed the celebrated Onsager solution\cite
{Onsager}, the dimer method\cite{Kasteleyn,Fisher,McCoy} fully exploits
the combinatorial and group theoretical properties of the lattices by
relating the partition function ${\cal Z}$ to the generating function ${\cal 
Z}_d$ of close--packed dimer configurations. Though the method is in
principle independent from the dimensionality of the lattice, the
corresponding analysis in three dimensions has never been developed, due to
the difficulties in extending Kasteleyn's Theorem on lattice orientation. In
this paper, we deal with the issue of generalizing the dimer approach to the
case of non-abelian $2D$ lattices of high topological genus which should in
fact be equivalent to higher dimensional lattices. Indeed the $3D$ cubic
lattice can be considered as a handlebody $2D$ lattice of genus $g=N/4$
where $N$ is the number of sites of the lattice. This hints that a non
vanishing ratio $g/N$ may be related to an effective dimension $D>2$ of the
lattice. 
The great difficulty of the problem  suggests
that a possible concrete way to analyze such lattices is to
consider at first graphs
possessing the largest possible symmetry group, and the best candidates
appear to be the group graphs already well--known in the mathematical
literature \cite{Cox1,Mag1,Mag2,Conway}.
In this framework we may consider the $2D$ planar lattice with periodic
boundary conditions as an example of abelian lattice group (translational
symmetry group) with genus $g=1$, whereas an example of finite non-abelian
lattice with $g=0$ is given by the fullerene--like lattice studied in ref. 
\cite{RR1}.

The paper is organized as follows. In Sec.{\bf II} we outline the basic
ideas concerning application of the dimer method to group lattices. In Sec.
{\bf III} we define an extended lattice group $H$ and relate it to the
homology and cohomology groups 
(${\rm mod} 2$) $H_1,H^1$ of the original lattice $\Lambda $, define the Grassmann
algebra over the decorated lattice $\Lambda ^{\#}$ and the Pfaffians as
function on orbits of $H\,^1.$ In Sec.{\bf IV} we apply the results of Sec.
{\bf III} to discrete groups of particular interest such as the Klein group 
$L(2,7)$  (of order $N=168$ and genus $g=3$) and discuss the role of external 
automorphisms. In Sec.{\bf V} we
analyze in detail the orbits of $H_1,H^1$ under the action of $H$, define an
invariant duality map $\varphi :H_1\rightarrow H^1$ and auxiliary functions
of relevant computational interest. Finally, in Sec.{\bf VI} we construct
explicitly the irreps of $H$, apply them to the harmonic analysis on $
\Lambda ^{\#}$ and derive the expansion for ${\cal Z}$. Few preliminary
numerical results are also given.

\section{Group lattices as Ising lattices}

We recall here key points of the dimer procedure which are specific to group
lattices (see also refs. \cite{LRR1}) :

{\bf (1)}. A discrete group $G$ is defined by a presentation given in terms of a
set of $p$ generators $A_k,\,\,k=1,...,p$ and $t$ relators $P_i
\;\;\;i=1,\dots ,t$. The $P_i$ are words in the generators equivalent to the 
identity. Let ${\cal F}$
be the free group on the $A_k\;\;k=1,\dots ,p$. Let ${\cal N}$ be the
minimal normal subgroup of ${\cal F}$ containing all relators. Then by
definition $G={\cal F}/{\cal N}$.

{\bf (2)}. The Cayley lattice $\Lambda $ for a group is defined by giving a
map $L:G\to {\bf R}^3$ where $L(g)\in {\bf R}^3$ is the point of ${\bf R}^3$
corresponding to the group element $g\in G$. A bond of color $k$ is then a
line joining $L(g)$ to $L(A_kg)$. The genus $g$ of $\Lambda \,$ is that of
the surface of minimal genus $S\subset {\bf R}^3$ on which $\Lambda $ can be
drawn. The Ising Hamiltonian is then defined as 
\begin{equation}
E=-\sum_{h\,\in G}\sum_{k=1}^pJ_k\,\,\sigma _h\,\sigma _{A_k\,h}\;\;\;,
\label{eq5}
\end{equation}
where the $\{\sigma _h=\pm 1\}$ are the spin variables and $\{J_k\}$ the
exchange interactions between connected spins.

{\bf (3)}. Each relator $P_i$ is then represented on $\Lambda $ as
a closed circuit $\zeta (P_i)$ made of oriented colored bonds. If $\zeta
(P_i)$ encloses a simply connected region (tile) on $S$ then $P_i$ is called
local relator. If $S$ has genus $g=0$ then all relators are local. Of
particular interest are models where $g$ is large.

{\bf (4)}. The group lattice $\Lambda $ is interesting on its own but cannot
be related directly to the partition function of an Ising model and to do this
we must consider\cite{McCoy} a decorated lattice $\Lambda ^{\#}$. This
amounts to replace each site of coordination $q$ $(q>2)$ of the original
lattice by a sublattice containing $3(q-2)$ points and $4q-9$ decorating
bonds. The Ising partition function is then related to the dimer covering
generating function on the decorated lattice.

{\bf (5)}. In order to compute the generating function we orient $\Lambda
^{\#}$ according to the Kasteleyn prescription by assigning arrows to each
bond  inherited from $\Lambda $ 
in such a way that 
for any closed circuit $\ell $ on $\Lambda ^{\#}$, the number
of bonds of ${\cal \ell }$ oriented clockwise is of opposite parity to the
number of sites enclosed by ${\cal \ell }$. For the decorating bonds see 
\cite{McCoy} or Sec.{\bf V}.

{\bf (6)}. The Kasteleyn rules define completely the orientation for
lattices of genus $g=0$, whereas for lattices of higher genus we have to
deal with further sign fixing for loops which are not homologically trivial,
i.e. not the boundary of a union of tiles. The assignment of arrows to 
$\Lambda ^{\#}$ (or $\Lambda$) is not invariant under the action of $G$ but
rather under an extension $H$ of $G$  closely related to the homology $H_1$ and
cohomology $H^1$ groups of $\Lambda $.

({\bf 7}). The dimer covering generating function of the lattice can be
expressed as a weighted sum of Pfaffians $P\!f(\phi )$ , where $\phi \in H^1$
and with sign given explicitly by the $H$--invariant function $\theta (\zeta
),\,\zeta \in H_1$ defined in Sec.{\bf V}. Harmonic analysis on 
$H$ allows to factorize Pfaffians into determinants of lesser order and external 
automorphisms induce identifications between Pfaffians.

\section{The extended lattice group}

In this section we first discuss the group extension $H$ of $G$. Next
we show that $H$ partitiones the homology $H_1$ and the cohomology $H^1$ groups of $\Lambda$
into non--intersecting orbits characterized in terms of sign functionals.
Their role in the expansion of the dimer generating
function is then analyzed. 

\subsection{The groups $H$, $H_1$ and $H^1$} 
The extended lattice group $H$ can be obtained from $G$ by replacing the
relators $P_i\,$with new relators containing the following elements:

\begin{itemize}
\item[({\bf i})]  If $P_i,P_j\in G$ are local, then $P_iP_j^{-1},\;P_i^2,
\;P_i\,A_kP_i^{-1}A_k^{-1}$ are relators in $H$;

\item[({\bf ii})]  For generic relators $P_i,P_j\in G$ then $P_i^2,\;P_j^2,
\;P_iP_jP_i^{-1}P_j^{-1}$ are relators in $H$.
\end{itemize}

We can write then $P_j =Q$ ($Q^2=1$) for all local relators $P_j$, with $
A_kQ=QA_k$, i.e. $Q$ is a central element that we call central signature.
Putting $Z_i=P_i Q$ the non local relators will be written as $
\;Z_i^2,\;Z_iZ_jZ_i^{-1}Z_j^{-1}$ where in general $A_kZ_i\neq Z_iA_k$ . The 
$\{Z_i\}$ generate an abelian normal subgroup $HZ \subset H$ . Particular
examples of this extension will be discussed further on.

Non trivial loops form a chain group $C_1(\Lambda ^{\#},{\bf Z_2)}$ . The
equivalence classes of $C_1(\Lambda ^{\#},{\bf Z}_2)$ modulo boundaries form
the homology group $H_1(\Lambda ^{\#},{\bf Z}_2)$. The class of
multiplicative functionals on $H_1(\Lambda ^{\#},{\bf Z}_2)$ with values $
\pm 1$ are then the elements of the cohomology group $H^1(\Lambda ^{\#},{\bf 
Z}_2)$. Notice that $H^1(\Lambda ^{\#},{\bf Z}_2)\sim H^1(\Lambda ,{\bf Z}_2)
$ and $H_1(\Lambda ^{\#},{\bf Z}_2)\sim H_1(\Lambda ,{\bf Z}_2)$ and
therefore we denote them briefly by the symbols $H^1$, $H_1$ respectively.
When dealing with elements of $HZ$, $H_1$ we may use addition instead of the
product as composition rule .We write $g\sim g^{\prime }$ if $g,g^{\prime
}\in H$ define the same site on $\Lambda .$

$H_1$ is isomorphic to $HZ$. To see it consider the closed chain $\ell $ on $
\Lambda $ as defined by the sequence $g_0=g,\;g_p=h_pg_{p-1},\;p=1,\dots ,n$
with $h_{n+p}=h_p$. $\zeta (\ell )$ is then defined as 
\begin{equation}
\zeta (\ell )=g_0^{-1}h_nh_{n-1}\dots h_1g_0=g_p^{-1}h_{p-1}h_{p-2}\dots
h_1h_nh_{n-1}..h_{p+1}g_p\;\;\;.  \label{eq30}
\end{equation}
Starting from $g_0$ we move $n$ steps across bonds in $\Lambda $ and each
bond defines an element $h_i\in H$, where $h_i=A_k^{\pm 1}$, $A_k$ being a
generator of $H$. If $g_n \sim g_0$, $\ell $ is closed, $\zeta (\ell
)\in HZ$ does not depend on the choice of $g_0$ on $\ell $ and defines a map 
$\zeta :C_1(\Lambda ,{\bf Z}_2)\to HZ$. 
Adding a boundary $\ell _0$ to $\ell$ (i.e. adding the boundary of a union of tiles
on the lattice) 
amounts to replace a sequence of bonds $h_{i_1},\dots ,h_{i_n}$ by an
equivalent one obtained by using local relators only , therefore $\zeta
(\ell +\ell _0)=\zeta (\ell )$ and  $\zeta \,$ induces the map $\kappa
:H_1\to HZ$. Given inversely an element $\zeta \in HZ$, expressed
in terms of relators as $\zeta =h_n\dots h_1,$ a corresponding closed chain $
\ell (\zeta )\in $ $C_1(\Lambda ,{\bf Z}_2)$, and therefore a cycle in $
H_1(\Lambda ,{\bf Z}_2)$, is given by a sequence of $g_i$, $i=0,\dots ,n$, with $
g_0={\bf 1}$ and $h_ig_{i-1}=g_i$. Also if $\zeta =h_n\dots h_1,\zeta
^{\prime }=h_{m+n}\dots h_{1+n}\,\in HZ$ and $g_n\sim g_{m+n}\sim {\bf 
1\,}$the chain $\ell (\zeta +\zeta ^{\prime })=\ell (\zeta )+\ell (\zeta
^{\prime })\in C_1(\Lambda ,{\bf Z}_2)$ defined by $g_0={\bf 1\,}$and $
h_ig_{i-1}=g_i, i=1...m+n$ corresponds to the element $\zeta \,+\zeta ^{\prime }
\in HZ$. Hence $\kappa $ is a group isomorphism.

Right multiplication on $\Lambda $ by an element $h\in H$ translates the
group lattice and  if we replace $g_i$ with $g_ih$ in $\ell $ we obtain a
closed path $\ell _h$ which is the right translation of $\ell $ by $h$ and a
corresponding element $\zeta (\ell _h)=h^{-1}\;\ell \;h$ which defines the
action of $H$ on $\ell \in HZ$. In this way $H$ acts naturally on $\Lambda
^{\#}$ , $H_1$ and $H^1$ and partitions $H^1,H_1$ into non--intersecting
orbits. The use of orbits greatly simplifies the computation of ${\cal Z}$.

\subsection{Sign functionals and lattice orientation}
We assign an orientation to $\Lambda ^{\#}$ according to the Kasteleyn rules
and to each site $h\in H$ a Grassman variable $a(h)$, with an anticommuting
wedge product $a(h)\wedge a(h^{\prime })=-a(h)\wedge a(h^{\prime })$.
Reversing the arrows on all bonds sharing the same site $i$ corresponds to
the change $a(h)\to -a(h)$.

Let $\ell \,$ given by (\ref{eq30}). Each $h_i$ is of the form $
A_{k_i}^{p_i}$ where $p_i=\pm 1$ determines the orientation of the arrow
in the bond. Closing $\ell $ means that we must identify $a(g_0)$ and $
a(g_n) $ as 
\begin{equation}
a(g_0)=p_na(g_n)  \label{eq50}
\end{equation}
We define then 
\begin{equation}
\eta (\zeta (\ell ))=-\prod_{i=1}^np_i \; \; \;.
\end{equation} 
If $\zeta $ is
trivial then $\eta (\zeta )=1$. All $\eta (\zeta )$ defined in this way,
hereafter called sign functionals , are characterized by a particular
recursion relation which can be proved as follows. 
Following ref.\cite{McCoy}
we introduce a reference dimer configuration ${\cal C}_0$ on $\Lambda ^{\#}$
consisting of all bonds of $\Lambda ^{\#}$ inherited from $\Lambda $. By
definition, any
other dimer configuration ${\cal C}$ when superposed on ${\cal C}_0$
generates transition cycles $\zeta \in H_1$. In this proof we use transition
cycles only. 
We consider representative chains $\ell, \ell',\ell'' \in C_1(\Lambda^\#,{\bf Z}_2)$
of the pair of intersecting cycles $\zeta ,\zeta
^{\prime }$ and of the cycle $\zeta +\zeta ^{\prime }$ as in Fig.1. 
Chains $\ell, \ell'$ run over the sequence of sites $P_i$, $i=0,1,2,7,8,9$
and $i\,=5,6,7,2,3,4$ respectively, while $\ell''$ runs
over $i=0,1,3,4$ and $i=5,6,8,9$ but not $i=2,7$. We denote with $B_{ij}$
the bonds joining $P_i,P_j$ and with $n\left( \ell \right)$ 
the number of
anticlockwise arrows on $\ell$. The key point is that bonds $B_{12},B_{13}$
are not contained in the intersection of $\zeta$ and $\zeta+\zeta^{\prime}$
and thus $B_{12},B_{13}\notin {\cal C}_0$ whereas $B_{01}\in {\cal C
}_0$. Similarly $B_{34},B_{27},B_{56},B_{89} \in {\cal C}_0$. By orienting 
$\Lambda ^{\#}$ according to the Kasteleyn rules and counting arrows we
see that $n\left(\ell''\right) \equiv n\left(\ell 
\right) +n\left( \ell' \right) +1 \, ({\rm mod} 2)$.
The general case where $B_{ij}$ are replaced by sequences of bonds
corresponds to the addition of boundaries to $\ell, \ell', \ell'' $
and leads to the same formula. Hence $n(\ell)$ is a function in
$H_1$ and $\eta (\zeta
+\zeta ^{\prime })=-\eta (\zeta )\eta (\zeta ^{\prime })$ if $\zeta ,\zeta
^{\prime }$ intersect and $\eta (\zeta +\zeta ^{\prime })=\eta (\zeta )\eta
(\zeta ^{\prime })$ otherwise , i.e. 
\begin{equation}
\eta (\zeta +\zeta ^{\prime })=\eta (\zeta )\eta (\zeta ^{\prime
})(-1)^{\Omega \left( \zeta ,\zeta ^{\prime }\right) }  \; \; \;,
\label{etabeta}
\end{equation}
where $\Omega \left( \zeta ,\zeta ^{\prime }\right) =0,1$ is the
intersection number $({\rm mod} 2)$ of $\zeta ,\zeta ^{\prime }$.
 Given a sign functional $\eta
(\zeta )$ then $\phi (\zeta )\eta (\zeta )$ , $\phi \in H^1$, is another
sign functional and so is the $H$--invariant function $\theta (\zeta
),\,\zeta \in H_1$ defined in Sec.{\bf V. }

\subsection{Pfaffians and dimer covering generating function}
Consider now the form 
\begin{equation}
f=\frac 12\sum_{h,h^{\prime }\in \Lambda ^{\#}}x_{h\,h^{\prime
}}\,a(h)\wedge a(h^{\prime })\;\;\;,  \label{eqform}
\end{equation}
where $x_{h\,h^{\prime }}=0$ if $h$ and $h^{\prime }$ are not connected in $\Lambda^\#$
(neighbors). The factor $\frac 12$ is inserted in order to avoid double
counting, the activity is given by $x_b=$ $x_{h\,h^{\prime
}}=-x_{h^{\prime }\,h}=\coth (\beta J_{h\,h^{\prime }})$ if the bond $b$ is
oriented with the arrow from $h$ to $h^{\prime }$, $\beta $ is the
inverse temperature and $J_{h\,h^{\prime }}$ the exchange interaction
between spins ($J_k$ in the previous notation) which depends only on the
color $k$ of the bond. Under the stated conditions decorating bonds have $
x_{h\,h^{\prime }}=1$. For the sole purpose of safe handling of signs we
assume initially $x_b>0$ . All arrows on $\Lambda ^{\#}$ and all signs in
(\ref{eqform}) are then fixed by the sign functional $\eta$. The Pfaffian $
P\!f(\eta )$ is defined by 
\begin{equation}
f^{M/2}=2^MM!P\!f(\eta )\;a(1)\wedge a(2)\wedge \dots \wedge a(M)\;\;\;.
\label{pfaffian}
\end{equation}
where $M=3(q-2)N$, $N$ is the order of the group (or lattice size) and $q$
is the lattice coordination. From (\ref{pfaffian}) one derives the
well-known relation 
\begin{equation}
P\!f(\eta )^2=Det(X(\eta ))  \label{det}
\end{equation}
where $X(\eta )$ is the $M\times M$ matrix of elements $x_{ij}$. In general $
P\!f(\eta )$ is multilinear function of the $x_b$ considered as independent
variables. The partition function ${\cal Z}$ is then given by
\begin{equation}
{\cal Z}={\cal Z}_d\,2^N\,\prod_a\sinh (\beta J_a)^{1/2}\;\;\;,  \label{Ztot}
\end{equation}
where $a$ runs on all the oriented bonds of the undecorated group lattice
and 
\begin{equation}
{\cal Z}_d=2^{-g}\sum_\eta s_\eta \,P\!f(\eta ) \; \; \; s_\eta=\pm 1 \;\;\;.  \label{Zd}
\end{equation}
To each dimer configuration ${\cal C}$ we associate the contribution of $
{\cal C}$ to $P\!f(\eta )$, a monomial $M({\cal C})=\prod_bx_b$ where $b$
runs over all oriented bonds of ${\cal C}$.
 Because of our conventions $M(
{\cal C})$ has sign $\eta (\zeta )$ where $\zeta$ is the superposition of 
${\cal C}$ and ${\cal C}_0$. In general $\zeta $ is the sum of non
intersecting cycles $\zeta =\sum_i \zeta_i$ so that $\eta (\zeta
)=\prod_i\eta (\zeta _i)$. The signs $s_\eta$ must then be chosen in such a way as
to set equal to $1$ in (\ref{Zd}) the coefficient of all $M({\cal 
C})$ appearing in ${\cal Z}_d$. Once this is done in Sec.{\bf V } the $x_b$
can be given any sign.

\section{$2D$ Ising lattice and the $L(2,7)$ lattice group}

The above group extension procedure can be naturally applyed to a wide class of lattice
groups. Here we discuss two examples: the abelian $2D$ Ising lattice (of genus $g=1$) and 
the $L(2,7)$ Klein group. The latter is non-abelian and of genus $g=3$, and its analysis 
should hopefully be of interest in the analysis of more general structures of dimension $D>2$.

\subsection{$2 D$ Ising lattice}

The Onsager solution for the $2D$ Ising lattice made use of a rectangular $
n\times m$ lattice with sites labelled by $i=1,\dots ,n$ and $j=1,\dots ,m$
( $n$ even). In this lattice we identify opposite sides, giving it a
toroidal ( $g=1$) topology and turning in a group lattice $G_{nm}$. This
last property makes it possible to apply harmonic analysis, i.e. Fourier
transform methods, which eventually lead to the final formula. The group $
G_{nm}$ of the lattice is defined by the presentation 
\begin{equation}
ST=TS,\;\;\;S^n={\bf 1},\;\;\;T^m={\bf 1}\;\;\;,  \label{presGnm}
\end{equation}
for all integers $n,m$.
However in order to satisfy the Kasteleyn rules we must use a central
extension $H_{nm}$ of (\ref{presGnm}) defined by: 
\begin{eqnarray}
ST &=&Q\,TS\;,\;\;\;S^n=Z_1\;,\;\;\;T^m=Z_2\;,  \nonumber \\
Q^2 &=&Z_1^2=Z_2^2=1  \nonumber \\
Q\,Z_1 &=&Z_1\,Q\;,\;\;\;Q\,Z_2=Z_2\,Q\;,\;\;\;Z_1Z_2=Z_2Z_1\;.
\end{eqnarray}
Sites are labeled by elements $h\in H_{nm}$ with the condition 
\begin{equation}
a(h\,Q)=-a(h)\;\;.  \label{condGnm}
\end{equation}
The $H_{nm}$ lattice is actually a covering of the $G_{nm}$ lattice with
corresponding variables $a(h)$ identified by the above condition which
embodies the Kasteleyn rules.

Besides this conditions we must set 
\begin{equation}
a(h\,Z_1)=\epsilon _1\,a(h)\;,\;\;\;a(h\,Z_2)=\epsilon _2\,a(h)\;,
\end{equation}
where $\epsilon _1$ and $\epsilon _2$ take the values $\pm 1$ in all the
possible combinations. These correspond to the $2^{2g}=4$ possible choices
of $\phi \in H^1$. It is tempting but misleading to label spin sites on the
lattice by elements in $G_{nm}$. In this case the orientation would no
longer be invariant under $G_{nm}$, a clear sign that the true symmetry is that
of the extended group $H_{nm}$.

\subsection{Non--abelian group lattices of genus $g>1$}
We examine now other types of symmetries which require non--central group
extensions and which are of interest for models in $D>2$. The natural
further step is provided by a vast array of discrete groups many of which
are discussed in detail in the literature (see \cite{Cox1,Mag1,Mag2,Conway}
). Of particular interest are the Klein groups $L(2,p)$, where $p$ is prime.
The non-abelian group of genus $g=0$ (the fullerene lattice) studied in ref. 
\cite{RR1} corresponds to $p=5$. The next interesting example which is at
the same time non abelian and has a non trivial topology is given by the
lattice group $L(2,7)$ (also called $T(2,3,7)$ in the context of hyperbolic
tessellations). The $L(2,7)$ group , briefly called $G$, has
order $168$ and is defined by the presentation 
\begin{equation}
U^7={\bf 1}\;,\;\;\;V^2={\bf 1}\;,\;\;\;(UV)^3={\bf 1}\;,\;\;\;(VU^3)^4={\bf 
1}\;.  \label{presG168}
\end{equation}
The analysis of a lattice possessing this kind of symmetry is one of the
chief results of this paper which hopefully opens the way to the
investigation of more general and interesting structures. The group lattice
given by (\ref{presG168}) has genus $g=3$ (produced by the non local relator 
$(VU^3)^4$) and can be tessellated by $24$ heptagons and $56$ hexagons, for
a total of $168$ bonds of type $U$ and $84$ bonds of type $V$ (see ref. \cite
{RaZe}, pag. 539--549, for more details on the lattice).

The group $G_{nm}$ given by (\ref{presGnm}) is abelian and has a non trivial
genus $g=1$. On a lattice of genus $g$ we expect that the Kasteleyn rule
determines the orientation up to $2g$ signs with a total of $2^{2g}$
configurations. Each close path on the lattice $\Lambda $ can be considered
as an element of the homology group $H_1$ and an assignment of all remaining
signs as an element of the cohomology group $H^1$. It is therefore sufficient to
assign signs on a suitable basis of $2g$ elements of $H_1$. In the toroidal
case this was done by giving $\epsilon_1$ , $\epsilon_2$. In fact $\,Z_1,Z_2$
are central in the group $G_{nm}$. 

The group $G$, as defined by (\ref
{presG168}), is non abelian and has a non trivial genus, in particular the $
Z_i$ are no longer central elements and we deal with the discussed non central
extension $H$ of $G$ defined by : 
\begin{eqnarray}
U^7 &=&Q\;,\;\;\;V^2=Q\;,\;\;\;(UV)^3=Q \; \; \;, \nonumber \\
W &\equiv &Q\,(VU^3)^4=\,Z({\bf 1})\;\;,  \label{presGext}
\end{eqnarray}
and additional relators which can be expressed in term of the auxiliary
elements 
\begin{equation}
Z(h)=h^{-1}W\,h,\;\;\;h\in H\;,
\end{equation}
characterized by 
\begin{equation}
Z(h_1)\,Z(h_2)=Z(h_2)\,Z(h_1)\;,\;\;\; \left(Z(h) \right)^2=1\;,\;\;\;\forall h_1,h_2,h\in
H\;.  \label{zid}
\end{equation}
The $Z(h)$ are not all independent and can be expressed in terms of the
subset of $2g=6$ elements 
\begin{equation}
Z_n\equiv Z(U^n)=U^{-n}Z({\bf 1})\,U^n\;\;.  \label{recur}
\end{equation}
Clearly $Z_n=Z_{n,({\rm mod} 7)}$ and $Z_0=Z({\bf 1})=Z_7$. The following
identities equivalent under (\ref{zid}) 
\begin{equation}
Z_5Z_1Z_4Z_0Z_3Z_6Z_2={\bf 1}\;,\;\;\;Z_1Z_3Z_5Z_0Z_2Z_4Z_6={\bf 1}\;\;,
\label{zident}
\end{equation}
reduces to $6$ the number of independent elements $Z_i$ and
can be proved by repeated application of relators (\ref{presGext}) but not
of (\ref{zid}). 

A generic $\zeta \in H_1\,$can be
always written in the additive form $\sum_in_iZ_i$ where $n_i=0,1.$ Because
of $\sum_{i=0}^6Z_i=0$ we can always choose the $n_i$ in such a way as to
have $\sum_in_i=0\,\,$or $1.$ All elements $Z(h),\;h\in H,$ can be obtained
by repeated conjugation of the $Z_n$ by $U$ and $V$. Conjugation by $U$ is
trivial, i.e. $U^{-1}Z_n\,U=Z_{n+1}$. Conjugation by $V$ is less obvious. By
using (\ref{presGext}) and not (\ref{zid}) we find that 
\begin{eqnarray}
V^{-1}Z_0V &=&Z_4\;,\;\;V^{-1}Z_1V=\,Z_3^{-1}Z_0^{-1}\;,\;
\;V^{-1}Z_2V=Z_4^{-1}Z_2^{-1}Z_0^{-1}\;,  \nonumber \\
V^{-1}Z_3V
&=&\,Z_4^{-1}Z_1^{-1}\;,V^{-1}Z_4V=Z_0\;,\;\;V^{-1}Z_5V=Z_5^{-1}\;,\;
\;V^{-1}Z_6V=Z_6^{-1}\;,  \label{conjV}
\end{eqnarray}
which can be further simplified by using (\ref{zid}). In the additive form ( 
\ref{conjV})and (\ref{recur}) can we written in full generality as: 
\begin{equation}
h^{-1}Z_ih=\sum_{k=0}^6P_{ik}(h)Z_k\,\,,\,\,\,h\in H  \label{repmod2}
\end{equation}
where $P(h)\,$is a representation ${\rm mod}\,2\,$of $H.$

Moreover the group $H$ has an external automorphism $\nu $ given by :

\begin{eqnarray}
\nu (V) &=&V^{-1}=-V  \nonumber \\
\nu (U) &=&U^{-1}  \nonumber \\
\nu (Z_p) &=&Z_{4-p}^{-1}=Z_{4-p}  \label{external}
\end{eqnarray}

\section{Orbits of homology and cohomology groups}
As anticipated, the analysis of orbits and of external automorphisms
plays a central role in the computation of ${\cal Z}_d$. We thus give here
the explicit construction of such orbits in $H_1$ and $H_1$, together with
their duality map.

\vskip .5cm
A functional $\phi \in H^1$ can be defined by the equivalent conditions:

\begin{equation}
a(gZ_i)=\epsilon _ia(g)\;,\phi (Z_i)=\varepsilon _i\;\,\,,\,\,i=0,\dots
,6\,\,\,\,\,\,\,\,;\,\,\,\,\,\,\prod_{i=0}^6\epsilon _i=1
\end{equation}
$\phi $ is then identified by $(\epsilon _0,\dots ,\epsilon _6)$ . Let $
\zeta \in H_1$ and let $h^{-1}\zeta h$ be the translated cycle. Then the
translated functional $\phi _h$ is defined by 
\begin{equation}
\phi _h(h^{-1}\zeta \,h)=\phi (\zeta )\,\,.  \label{funch}
\end{equation}
We have then: 
\begin{equation}
\phi _U(U^{-1}Z_iU)=\phi (Z_i)=\epsilon _i=\phi _U(Z_{i+1})\,\,\,,
\label{phiU}
\end{equation}
and hence 
\begin{equation}
\phi _U(Z_i)=\epsilon _{i-1}\;,\;\;\;\phi _{U_{-1}}(Z_i)=\epsilon
_{i+1}\;\;\;.  \label{phiUep}
\end{equation}
Similarly we find 
\begin{equation}
\phi _V(V^{-1}Z_0V)=\phi (Z_0)=\epsilon _0=\phi _V(Z_4)\;\;\;,  \label{phiV0}
\end{equation}
and 
\begin{eqnarray}
\phi _V(Z_0) &=&\epsilon _4\;,\;\;\;\phi _V(Z_3)=\epsilon _1\epsilon
_4\;,\;\;\;\phi _V(Z_2)=\epsilon _0\epsilon _4\epsilon _2\;,  \nonumber \\
\phi _V(Z_1) &=&\epsilon _3\epsilon _0\;,\;\;\;\phi _V(Z_5)=\epsilon
_5\;,\;\;\;\phi _V(Z_6)=\epsilon _6\;\;\;.  \label{phiVn}
\end{eqnarray}
There are $2^{2g}=64$ ($g=3$ denoting the genus of $\Lambda $) different
elements in $H^1$ which fall in $5$ different orbits containing $1,7,7,21,28$
elements . We list these orbits with only one signature out of each
cyclically permuted septet:

\begin{eqnarray}
A &:&\;\;(1,1,1,1,1,1,1)\;\;\;\;\;\;({\rm trivial})  \nonumber \\
B &:&\;\;(1,1,-1,1,-1,-1,-1)  \nonumber \\
C &:&\;\;(-1,-1,-1,1,-1,1,1)  \nonumber \\
D &:&\;\;(-1,-1,-1,1,-1,-1,-1)\;,(-1,-1,1,1,1,-1,-1)\;,\;(1,1,-1,1,-1,1,1) 
\nonumber \\
E &:&\;\;(-1,1,-1,1,-1,1,-1),\;(1,-1,-1,1,-1,-1,1),\;(-1,1,1,1,1,1,-1), 
\nonumber \\
&&\ \ \,\,\,(1,-1,1,1,1,-1,1)  \label{orbfunct}
\end{eqnarray}

The orbits $B,C\,$ are mapped into each other by the action of $\nu$. This
proves incidentally that $\nu $ is external. We also list in similar fashion
the dual orbits in $H_1:$

\begin{eqnarray}
A &:&\;\;(0,0,0,0,0,0,0)\;\;\;\;\;\;({\rm trivial})  \nonumber \\
B &:&\;\;(1,1,0,0,1,0,1)  \nonumber \\
C &:&\;\;(1,0,1,0,0,1,1)  \nonumber \\
D &:&\;\;(0,0,1,0,1,0,0)\;,(0,1,1,0,1,1,0)\;,\;(0,1,0,0,0,1,0)  \nonumber \\
E &:&\;\;(1,0,0,0,0,0,1),\;(1,1,1,0,1,1,1),\;(1,1,0,0,0,1,1),\;  \nonumber \\
&&\,\,\,\,\,(1,0,1,0,1,0,1)  \label{signorbits}
\end{eqnarray}
as it can be checked by using (\ref{phiV0})--(\ref{phiVn}).

\noindent
For generic elements $\zeta ,\zeta ^{\prime }\,\in \,H_1$ we use the
additive form: 
\begin{equation}
\zeta =\sum_{i=0}^6 n_i Z_i\,\,\,\,\, , \,\,\,\,\,\zeta ^{\prime
}=\sum_{i=0}^6 m_i Z_i \; \; \;, 
\end{equation}
and define the intersection number $({\rm mod} 2)$
\begin{equation}
\tau (\zeta ,\zeta ^{\prime })=\tau (\zeta ^{\prime },\zeta )=(-1)^{\Omega
\left( \zeta ,\zeta ^{\prime }\right) } \; \;\;. 
\label{inter}
\end{equation}
$\Omega $ is the ${\bf Z}_2$ valued form of eq.(\ref{etabeta}) explicitly given by: 
\begin{equation}
\Omega \left( \zeta ,\zeta ^{\prime }\right) =\sum_{i,k=0}^6\chi
(i-k)\,n_i\,m_k 
\end{equation}
with $\chi (i)=1$ if $i\equiv 1,2,5,6 \, ({\rm mod} 7)$ and $\chi (i)=0$ otherwise,
as it can be verified on the graph $\Lambda ^{\#}$.
$\tau (\zeta ,\zeta ^{\prime })$ is invariant under conjugations by $H$,
see (\ref{conjV}), and is multiplicative, i.e. 
\begin{equation}
\tau (\zeta ,\zeta ^{\prime }+\zeta ^{\prime \prime })=\tau (\zeta ,\zeta
^{\prime })\,\tau (\zeta ,\zeta ^{\prime \prime })\,\,\,.  \label{tauxtau}
\end{equation}
We define also: 
\begin{equation}
\theta (\zeta )=(-1)^{\frac 12\sum_{i,k=0}^6\chi
(i-k)\,n_i\,n_k}\,\,\,\,\,\,\,\,\,\,,  \label{theta}
\end{equation}
so that $\theta (Z_i)=\theta ({\bf 1})=1$ and $\theta (\zeta )$ is a sign
functional as $\eta$ in (\ref{etabeta}): 
\begin{equation}
\theta (\zeta +\zeta ^{\prime })=\theta (\zeta )\theta (\zeta ^{\prime
})\tau \left( \zeta ,\zeta ^{\prime }\right) \,\,\,\,.  \label{thetaxtheta}
\end{equation}
All these definitions can be extended naturally to a wide class of lattice
groups.

Let then : 
\begin{equation}
m_i=\sum_{k=0}^6\chi (i-k)\,n_k  \label{metrica}
\end{equation}
and define the duality map $\varphi \,\,,H_1\rightarrow H^1$:$\,\,$ 
\begin{equation}
\varphi \left( \sum_{i=0}^6n_iZ_i\right) =\left(
(-1)^{m_0},....,(-1)^{m_6}\right)  \label{themap}
\end{equation}
The $2^{2g}$ elements of $H_1$,$H^1$ are labeled as $\zeta _I,\phi _I,$where 
$\varphi \left( \zeta _I\right) =\phi _I$ , by an index $I=1,...,2^{2g}$
with $\zeta _1,\phi _1$ the trivial elements and sorted in such a way that
in (\ref{signorbits}),(\ref{orbfunct}) $\varphi$ maps corresponding orbits
in $H_1\,,H^1$. Clearly then $\phi _I\left( \zeta _K\right) =\tau (\zeta
_I,\zeta _K)$. We set $\eta _I\left( \zeta \right) =\phi _I\left( \zeta
\right) \theta \left( \zeta \right)$ , $s_I=s_{\eta _I}$ so that we may
label Pfaffians equally well with elements $\phi \in H^1$ and rewrite (\ref
{Zd}) as: 
\begin{equation}
{\cal Z}_d=2^{-g}\sum_I s_I P\!f(\phi _I) \; \; \; .
  \label{Zd1}
\end{equation}

The importance of orbits should be clear once we realize that in (\ref{Zd1})
for elements $\phi _I,\phi _J$ in the same orbit we have $P\!f(\phi
_I)=P\!f(\phi _J)$. In this way the effective number of different terms in $
{\cal Z}_d$ reduces to the number of orbits in $H^1$.

 Further reductions
arise from external automorphisms. When omitted, as in (\ref{Zd1}), we assume
summation ranges on $I,K$ to be $1...2^{2g}$. Since $\varphi$
commutes with the group operations it maps orbits in $H_1$ into dual orbits
in $H^1$ and we use the same label $A...E$ for pairs of dual orbits.
Moreover the $2^{2g}\times 2^{2g}$ matrix $\Phi=\Phi^T$ of elements $
\Phi _{KI}=2^{-g}\tau (\zeta _I,\zeta _K)$ is orthogonal. We have first of
all: 
\begin{equation}
\sum_I\Phi _{KI}^2=2^{-2g}\sum_I1=1 \; \; \; .
\end{equation}
We have next: 
\begin{eqnarray}
p(K,K^{\prime }) &=&\sum_I\Phi _{KI}\Phi _{K^{\prime }I}=2^{-2g}\sum_I\phi
_K(\zeta _I)\phi _{K^{\prime }}(\zeta _I)=2^{-2g}\sum_I\phi _K(\zeta _I+\xi
)\phi _{K^{\prime }}(\zeta _I+\xi ) \\
\ &=&2^{-2g}\sum_I\phi _K(\zeta _I)\phi _{K^{\prime }}(\zeta _I)\phi _K(\xi
)\phi _{K^{\prime }}(\xi )=p(K,K^{\prime })\phi _K(\xi )\phi _{K^{\prime
}}(\xi )  \nonumber
\end{eqnarray}
therefore $p(K,K^{\prime })=0\,$ if $\exists \xi :$ $\phi _K(\xi )\phi
_{K^{\prime }}(\xi )\neq 1$. But if $\forall \xi :\phi _K(\xi )\phi
_{K^{\prime }}(\xi )=1$ then $K=K^{\prime }$ therefore $p(K,K^{\prime })=0$
if $\,K\neq K^{\prime }$. This proves that $\Phi $ is orthogonal and that $
\Phi =\Phi ^{-1}$. 

We conclude this section with some rather technical  
formulas whose r\^ole will be crucial in
what follows (sect. {\bf VIII}).

The sum $\sum_K\theta (\zeta _K)$ can be readily evaluated by using a
standard basis for $H_1$ given by $X_i,Y_i\;,\;i=1...g$ such that

\begin{equation}
\theta (\sum_i^g\left( p_iX_i+q_iY_i\right)
\,)\,=(-1)^{\sum_{i=1}^g\,p_i\,q_i}\,\,
\end{equation}

where for instance:

\begin{equation}
X_1=Z_0,Y_1=Z_1,X_2=Z_1+Z_5,Y_2=Z_0+Z_2+Z_5,X_3=Z_0+Z_3,Y_3=Z_1+Z_3+Z_6
\end{equation}
In this form $\sum_K\theta (\zeta _K)$ factors into $g$ independent and
equal sums each yielding a factor $2$ and hence $\sum_K\theta (\zeta _K)=2^g$
.

Furthermore: 
\begin{equation}
2^{-g}\sum_K\tau (\zeta _I,\zeta _K)\theta (\zeta _K)=2^{-g}\,\theta (\zeta
_I)\sum_K\theta (\zeta _K+\zeta _I)\,\,=\,\theta (\zeta _I)\,\,\,\,.
\label{craxi}
\end{equation}
since $\zeta _K+\zeta _I$ runs over the whole group $H_1$ taking every
element once just as $\zeta _K$ and therefore $\sum_K\theta (\zeta _K+\zeta
_I)=\sum_K\theta (\zeta _K)=2^g$ $.$

\section{Irreps of H}

The group $H$ has $168\times 64=10752$ elements and in order to perform
harmonic analysis and obtain partial block diagonalization of $X(\phi )$ and
of Pfaffians we must find the unitary irreducible representations (irreps)
of $H$.

The trivial functional $(1,1,1,1,1,1,1)$ must be dealt with separately and
requires the construction of the irreps of the factor group $H_0$ of $H$ and
central extension of $G$ defined by 
\begin{equation}
U^7=Q\;,\;\;\;V^2=Q\;,\;\;\;(UV)^3=Q\;,\;\;\;(VU^3)^4=Q\;,  \label{presH0}
\end{equation}
(i.e. $Z_i={\bf 1}$) of order $168\times 2=336$. Therefore we have the
relators : 
\begin{equation}
a(gZ_i)=a(g)\;,\;\;\;a(gQ)=-a(g)\;\;\;.  \label{relH0}
\end{equation}
Because of (\ref{relH0}), only a subset of the unitary irreps of $H_0$ is
actually used in the harmonic analysis. In order to see it let us write such
irreps as $D_{\alpha \beta }^J(g)$ with $J$ a convenient label and $\alpha
,\beta =1,\dots ,d_J$ where $d_J$ is the dimension of the irrep . The matrix
elements satisfy the orthogonality relations: 
\begin{equation}
\sum_{g\in H_0}D_{\alpha \beta }^J(g)^{*}D_{\alpha ^{\prime }\beta ^{\prime
}}^{J^{\prime }}(g)=\delta ^{JJ^{\prime }}\delta _{\alpha \alpha ^{\prime
}}\delta _{\beta \beta ^{\prime }}\;\;\;.  \label{ort}
\end{equation}
Let us define 
\begin{equation}
a_{\alpha \beta }^J=\sum_{g\in H_0}a(g)D_{\alpha \beta }^J(g)\;\;\;.
\label{aj}
\end{equation}
Applying (\ref{aj}) to (\ref{relH0}) we get 
\begin{equation}
\sum_{g\in H_0}a(gQ)D_{\alpha \beta }^J(g)=-a_{\alpha \gamma }^J=a_{\alpha
\gamma }^JD_{\gamma \beta }^J(Q)\;\;\;.  \label{ztrans}
\end{equation}
But $Q$ is a central element and $Q^2={\bf 1}$. Therefore by Schur lemma $
D_{\gamma \beta }^J(Q)$ must be proportional to the identity, i.e. 
\begin{equation}
D_{\gamma \beta }^J(Q)=\kappa \delta _{\gamma \beta }\;,\kappa =\pm 1\;\;\;,
\label{diag}
\end{equation}
therefore from (\ref{ztrans}) 
\begin{equation}
\kappa \,a_{\alpha \gamma }^J=-a_{\alpha \gamma }^J\;\;\;,
\end{equation}
i.e. $\kappa =-1$. It follows that only irreps having $\kappa =-1$ actually
contribute to the Fourier expansion of $a(g)$. Those irreps which are also
irreps of the original group $G$ (i.e. those with $\kappa =1$) are absent
from the expansion. There are only $5$ irreps of $H_0$ with $\kappa =-1$ of
dimension $4,4,6,6,8$, satisfying separately the Burnside condition $
4^2+4^2+6^2+6^2+8^2=168$. The detailed form is listed in Appendix {\bf A}.
The matrix elements of these irreps are polynomials in $K=\exp (i\pi /7)$
with $K^7=-1$. By abuse of language we may write 
\begin{equation}
Q=-1\;,\;\;\;U^7=-1\;,\;\;\;V^2=-1\;,\;\;\;(UV)^3=-1\;,\;\;\;(VU^3)^4=-1\;,
\end{equation}
instead of (\ref{presH0}).

As for the remaining irreps of $H$ the most convenient way is to obtain them
as induced representations on the cosets of the subgroup $L$ of $H$
generated by: 
\begin{equation}
V\;\;,\;\;U^{-1}VU^2\;,\;\;\;U^{-6}VU^3\;,\;\;\;U^{-4}VU^4\;,\;\;
\;U^{-5}VU^5\;\;,  \label{genL}
\end{equation}
as well as the cosets of the group $\nu (L)$ generated by : 
\begin{equation}
V^{-1},\ \,\,\,\,UV^{-1}U^{-2},\ \,\,\,U^6V^{-1}U^{-3},\
\,\,\,\,U^4V^{-1}U^{-4},\ \,\,\,\,\,U^5V^{-1}U^{-5}.  \label{geniL}
\end{equation}
All generators and their inverses are of the form $U^{a(b)}VU^b\;,\;\;b=0,
\dots ,6$. Let set briefly $v=V$ and $t=U^{-1}VU^2$; we find then 
\begin{eqnarray}
U^{-6}VU^3 &=&v\,t\;,\;\;\;U^{-4}VU^4=Z_1(\,t\,v\,)^2=(\,t\,v\,)^2Z_4 
\nonumber \\
U^{-5}VU^5 &=&v\,t\,(\,v\,t^{-1})^2Z_1=Z_2Z_6v\,t\,(\,v\,\,t^{-1})^2\; 
\nonumber \\
(\,v\,\,t\,)^4
&=&Q\,Z_3\,\,\,\,\;,\,\,\,\,\,(\,t\,\,v\,)^4=\,Q\,Z_{1\,}Z_4\,\,\,\,\,.
\label{genLII}
\end{eqnarray}
It can be verified that $\left\{ v,\;t,\;Z_i,\;Q\right\} $ generate the
whole subgroup $L$. Let suppose that a representation $\lambda :L\to {\rm Hom
}({\cal L})$ is given on a linear space ${\cal L}$ of dimension $n$. Then $
\lambda $ can be extended to the induced representation $\mu :$ $H\to {\rm 
Hom}({\cal H})$ where ${\cal H}=\bigoplus_{p=0}^6{\cal L}_p$ and ${\cal L}_p$
are isomorphic to ${\cal L}$. Let $\psi _i$ be a basis on ${\cal L}$, $\psi
_{i,p}$ a basis on ${\cal L}_p$ and $\psi _{i,p+7}=-\psi _{i,p}\;$. The
action of $U$ on ${\cal H}$ is then defined by 
\begin{equation}
U\psi _{i,p}=\psi _{i,p+1}\;\;\;,  \label{Uaction}
\end{equation}
hence $U^7=-1$. Moreover we set 
\begin{equation}
w\psi _{i,0}=\psi _{k,p}\lambda _{k,i}(w)\;,\;\;\;w\in L\;\;\;.
\label{Laction}
\end{equation}
From (\ref{genLII})--(\ref{Laction}) derive the representation of $V$ on $
{\cal H}$ as follows: 
\begin{eqnarray}
V\psi _{i,b} &=&VU^b\psi _{i,0}=U^{a(b)}U^{-a(b)}VU^b\psi _{i,0}=  \nonumber
\\
U^{a(b)}\psi _{k,0}\lambda (U^{-a(b)}VU^b)_{k,i} &=&\psi _{k,a(b)}\lambda
(U^{-a(b)}VU^b)_{k,i}\;.  \label{Vrep}
\end{eqnarray}
Since $b$ can take all values $0,\dots ,6$ we deduce the complete
representation of $V$ on ${\cal H}$. The problem reduces now to that of
finding a suitable representation of $L$ on ${\cal L}$. Should we set $Q=Z_i=
{\bf 1}$, $L$ becomes the octahedral group $T(2,3,4)$ of order $24$. We are
interested in irreps of $L$ in which all $\lambda (Z_i)$ are diagonal and $
\lambda (Z_i^2)={\bf 1}$. Each $\lambda (Z_i)$ is then a diagonal block
matrix. In fact we have 
\begin{equation}
Z_0\psi _{i,p}=Z_0U^p\psi _{i,0}=U^pZ_p\psi _{i,0}=U^p\lambda (Z_p)\psi
_{i,0}=\lambda (Z_p)\psi _{i,p}\;\;\;.  \label{Zaction}
\end{equation}
It is then clear that $\mu ($ $Z_s)$ is the diagonal matrix with entries $
\lambda (Z_s)$, $\lambda (Z_{s+1})$,$\dots $, $\lambda (Z_{s+6})$ , where $
\lambda (Z_s)$ has $n$ eigenvalues $\xi _{i,s}=\pm 1,i=1,\dots
,n\;,\;\;s=0,\dots ,6$. It follows also that each vector $\psi _{i,p}$ has
eigenvalues 
\begin{equation}
Z_s\psi _{i,p}=\xi _{i,s+p}\psi _{i,p}\;\;\;,  \label{eigenpsi}
\end{equation}
which define the element $\Xi \in H^1:$ 
\begin{equation}
\Xi =(\xi _{i,p},\xi _{i,p+1},\dots ,\xi _{i,p+6})  \label{signsymbol}
\end{equation}

For the group $\nu (L)$ the corresponding formulae are obtained by setting $
v^{\prime }\,=V^{-1},\,\,\,t^{\prime }=UV^{-1}U^{-2}\,.$ We have then

\begin{eqnarray}
U^6V^{-1}U^{-3} &=&v^{\prime }t^{\prime
}\,\,\,\,,\,\,\,U^4V^{-1}U^{-4}=\,\,Z_3\,(t^{\prime }v^{\prime
})^2=(t^{\prime }v^{\prime })^2\,\,Z_0\,\,, \\
\,(v^{\prime }t^{\prime })^4 &=&Q\,Z_0^{-1},(t^{\prime }v^{\prime
})^4=Q\,Z_0^{-1}Z_3^{-1}\,\,.  
\label{extG}
\end{eqnarray}

If an irrep of $H$ contains a signature $\Xi $ it contains also the whole
orbit including its cyclical permutations. Non trivial functionals have $7$
distinct cyclical permutation of any $\Xi $ corresponding to $p=0,\dots ,6$
in (\ref{signsymbol}). The irrep $\mu $ of $H$ generated by $\lambda $ has
dimension $7n$ and as many signatures. If it contains different orbits then
is reducible. Therefore irreps can be grouped in disjoint subsets
characterized by orbits. Irreps belonging to orbits $A,D,E$ can be obtained
either from $L\,$or $\nu (L)$ whereas $B$ and $C$ can be obtained only from $
L\,\,$and $\nu (L)$ respectively. We list in appendix {\bf B} the explicit
representations $\lambda $ of the group $L$ corresponding to each orbit by
giving $\lambda (v)\,,\,\lambda (t)\,$ and $\lambda (Z_1)$. In this way we
obtain the irreps of the orbits $A,B,D,E$. Irreps of orbit $C$ can be
obtained by applying $\nu $ to irreps of $B$. Clearly orbits may appear in
some irreps with multiplicities $m=1,2,4$ where $n/m$ is the number of
inequivalent signatures under cyclical permutations contained in the orbit.
The method applied to orbits $A$ yields representations which reduce into
the irreps already discussed. The dimensions of the irreps satisfy a
separate Burnside condition 
\begin{equation}
\sum_{J\in O}d_J^2=n(O)N  \label{burnside}
\end{equation}
where $n(O)$ is the number of signatures belonging to a given orbit $O$.

\section{The decorated lattice and the partition function}

We need now a more explicit description of the decorated lattice $\Lambda
^{\#}$ along lines already discussed in ref. \cite{RR1}. Each element $h\in
H $ \thinspace \thinspace identifies uniquely a site of $\Lambda $ which we
also label with $h,$being intended that $hQ\sim hZ_i\sim Z_ih$ identify the
same site of $\Lambda $. $h$ is connected to other 3 sites $V\,h,Uh,U^{-1}h$
and is replaced in $\Lambda ^{\#}$ by $3$ sites $h_a,h_b,h_c$ to which we
associate Grasmann variables $a(h),b(h),c(h)$. The original oriented bonds $
h\rightarrow Vh,h\rightarrow Uh$ are replaced by oriented bonds $
h_a\rightarrow Vh_a,h_b\rightarrow Uh_c$ (with exchange interaction $
J_V,J_U\,$ respectively) together with the additional decorating bonds $
\,h_a\rightarrow h_c,h_c\rightarrow h_b,h_b\rightarrow h_a$. For each
Pfaffian we need boundary conditions of the kind discussed above: 
\begin{equation}
a(hZ_i)=\epsilon _ia(h)\,\,,\,\,a(Qh)=a(hQ)=-a(h)  \label{boundary}
\end{equation}
and the corresponding conditions obtained by replacing $a$ with $b,c$. Under
these assumptions the 2-form (\ref{eqform}) can be rewritten explicitly as : 
\begin{equation}
f=\sum_{h\in \Lambda }\left( a(h)\wedge c(h)+c(h)\wedge b(h)+b(h)\wedge a(h)+
\frac y2\,a(h)\wedge a(Vh)+x\,b(h)\wedge c(Uh)\right)  
\label{eqformdec}
\end{equation}
where $y=\coth (\beta J_V),$ $x=\coth (\beta J_U)$ .

Harmonic analysis on $\Lambda $ can be performed by recalling the Fourier
components of $a(h),b(h),c(h)$ given by $a(h)=\sum_{\alpha \beta
}^JD_{\alpha \beta }^J(h)^{*}a_{\alpha \beta }^J$, etc. . To this purpose we
consider the matrices ${\bf a}^J,{\bf b}^J,{\bf c}^J\,$ of elements $
a_{\alpha \beta }^J\,,b_{\alpha \beta }^J\,,c_{\alpha \beta }^J$ where $
\alpha ,\beta =$ $1...d_{J\text{ }}$. By using the orthonormality relations
(\ref{ort}) satisfied by the matrices $D_{\alpha \beta }^J(h)$ we rewrite $
f=\sum_Jf^J$ where 
\begin{equation}
f^J=Tr\left( {\bf a}^{J\dagger }\wedge {\bf c}^J+{\bf c}^{J\dagger }\wedge 
{\bf b}^J+{\bf b}^{J\dagger }\wedge {\bf a}^J+\frac y2\,{\bf a}^{J\dagger
}\wedge D^J(V)\,{\bf a}^J+x\,\,\,{\bf b}^{J\dagger }\wedge D^J(U){\bf \,c}
^J\right)  \label{formJ}
\end{equation}
and $J^{\dagger }$ labels the complex conjugate irrep of $J$. The key point
is that different pairs $J,J^{\dagger }$ lead to disjoint sets of Grassman
variables, moreover $f^J+f^{J\dagger }$ separates into the sum of partial
forms $\sum_\alpha (f^J)_{\alpha \alpha }+c.c.$~. Therefore the final
Pfaffian is the product of $d_J$ partial (and identical)
Pfaffians, which must be computed explicitly by use of (\ref{det}).
Once this is done we must determine the coefficients $s_I$ appearing in (\ref
{Zd1}). Since all the $s_I$ and Pfaffians $P\!f(\phi _I)$ sharing the same
orbit $O\,$of $\phi _I\,$are equal\thinspace we can set $S_O=n(O)\,s_I\;,P
\!f(\phi _I)=p\!f(O)\,,\,\forall \phi _I\in O$ and (\ref{Zd1}) reduces to: 
\begin{equation}
{\cal Z}_d=2^{-g}\sum_OS_O\,p\!f(O) \; \; \;.
 \label{ZdO}
\end{equation}

The sign in $P\!f(\phi _I)$ of a term associated with the path $\zeta $ is
given by $\eta _I(\zeta )=\tau (\zeta _I,\zeta )\,\theta (\zeta )$. The $
s_I\,$ must be chosen in such a way as to set the final coefficient of $
\zeta _K\,$ in the expansion (\ref{Zd1}) : 
\begin{equation}
2^{-g}\sum_Is_I\tau (\zeta _I,\zeta _K)\,\theta (\zeta _K)
\end{equation}
equal to $1$. Taking into account(\ref{craxi}) we see that the condition 
\begin{equation}
\forall I:2^{-g}\sum_Is_I\,\,\tau (\zeta _I,\zeta _K)\,=\sum_Is_I\,\Phi
_{IK}=\theta (\zeta _K)
\end{equation}
has the simple solution: 
\begin{equation}
s_I=\,\theta (\zeta _I)\,\,,
\end{equation}
and thus $S_A\;=1,S_B=S_C=7,S_D=21,S_E=-28\,$ and $S_A+S_B+S_C+S_D+S_E=8=2^g$.

\section{Pfaffians and harmonic analysis on the lattice}

The boundary conditions (\ref{boundary}) on the Grassmann variables $
a(h),b(h),c(h)$ imply strong restrictions on the Fourier components $
a_{\alpha \beta }^J\,\,,\,b_{\alpha \beta }^J\,\,,\,c_{\alpha \beta }^J$
similar to those already derived for $Q\,$. Consider now $P\!f\left( \phi
\right) $ where $\phi =\left( \epsilon _0,\epsilon _1,\epsilon _2,\epsilon
_3,\epsilon _4,\epsilon _5,\epsilon _6\right) $. From (\ref{boundary}) we
get~: 
\begin{equation}
\sum_{h\in H}a(hZ_i)D_{\alpha \beta }^J(h)=\epsilon _i\,a_{\alpha \gamma
}^J=a_{\alpha \gamma }^JD_{\gamma \beta }^J(Z_i)\;\;
\end{equation}
In all the explicit representation of $H$ listed in the Appendix {\bf B} the
matrices $D_{\gamma \beta }^J(Z_i)$ are diagonal and 
\begin{equation}
D_{\gamma \beta }^J(Z_i)=\epsilon _i^J(\gamma )\delta _{\gamma \beta
}\,\,\,\,\,.
\end{equation}
Clearly $a_{\alpha \gamma }^J,b_{\alpha \gamma }^J,c_{\alpha \gamma }^J=0$
unless $\epsilon _i^J(\gamma )=\epsilon _i\,,\;i=0...6.$ The orbit $O$ of $
\phi $ identifies a subset of irreps and in (\ref{formJ}) only a rectangular
submatrix of ${\bf a}^J,{\bf b}^J,{\bf c}^J$ of dimension $d_J\times
d_J\,/n(O)$ survives where $d_J$ is the dimension of the irrep and $
d_J\,/n(O)\,$the number of times a given signature in $O\,$ appears in the
irrep. The total number of surviving components of $a_{\alpha \gamma }^J\,$
among all irreps is in any case equal to the order $N$ of the group $G$,
i.e. $168$.

Each ${\bf a}^J,{\bf b}^J,{\bf c}^J$ is therefore partitioned into $
\,d_J\,/n(O)$ rectangular submatrices labeled by the signatures in $O$. Only
one of these matrices needs to be used because they all yield the same final
expression for the Pfaffian. Since in (\ref{formJ}) all partial forms $
f_{J\alpha }$ yield identical Pfaffians we may consider just one
contribution and drop the index $\alpha $. In this case the square of the
partial Pfaffian can be written as the determinant of the block matrix $
\Delta _J$ 
\begin{equation}
\Delta _J=\left[ 
\begin{array}{ccc}
y\,D^J(V) & {\bf 1} & -{\bf 1} \\ 
-{\bf 1} & 0 & {\bf 1-\,}x\,D^J(U) \\ 
{\bf 1} & -{\bf 1+\,}x\,D^J(U^{-1}) & 0
\end{array}
\right]  \label{delta}
\end{equation}
Notice that $\Delta _J^{\dagger }=-\Delta _J.$ Upon multiplication of the
last group of columns by $\left( {\bf 1-\,}x\,D^J(U)\right) ^{-1}$ and the
last group of rows by $\left( {\bf 1-\,}x\,D^J(U^{-1})\right) ^{-1}$ we find
as in \cite{RR1} : 
\begin{equation}
Det\left( \Delta _J\right) =Det\left[ y\left( {\bf 1-\,}x\,D^J(U)\right)
D^J(V)\,\,\left( {\bf 1-\,}x\,D^J(U^{-1})\right) +x\,\left(
D^J(U)-D^J(U^{-1})\right) \right]
\end{equation}
and therefore the effective dimension of the final determinant reduces to $
d_J$. The dimer generating function ${\cal Z}_d$ is then: 
\begin{equation}
{\cal Z}_d=2^{-g}\sum_OS_O\,p\!f(O)=2^{-g}\sum_OS_O\prod_{J\in
O}P\!f_J^{d_J\,/\,n(O)\,} \; \; \; .
\end{equation}
The degree of $P\!f_J$ in $x,y\,$ is $d_J,d_J/2\,$ and therefore the total
degree of $\,p\!f(O)$ in $x$ is $\sum_{J\in O}d_J^2/n(O)=\;N$ because of the
Burnside condition and half as much for $y.$ Not all $Det\left( \Delta
_J\right) \,$ are distinct, besides $Det\left( \Delta _J\right) =Det\left(
\Delta _{J^{\dagger }}\right) $ the external automorphism $\nu $ maps $
pf(B)\,$into $pf(C)\,$and viceversa thus $pf(C)=pf(B)\,$. The number of
independent Pfaffians is therefore further reduced to four $(64 \to 5 \to4$) by the existence of
external automorphisms.

If $d_J\,/\,n(O)\,$ is odd the sign of $P\!f_J$ $=\pm \sqrt{Det\left( \Delta
_J\right) }$ must be fixed unambiguously by computing it for instance in the
limit $\beta \rightarrow 0.$ ${\cal Z}_d\,$ becomes then a polynomial in $
x,y $ of degree $N,N/2$ \cite{note1}. 
In Fig.2 we give the plot of the specific heat corresponding to 
a ferromagnetic choice of the exchange interactions, $(J_U=J_V=1)$. 
The specific heat presents a very sharp peak which
prefigures the transition form and ordered ferromagnetic phase to a disordered
paramagnetic one. When $J_U,J_V$ have different signs or are both
antiferromagnetic (i.e. negative) the model is totally frustrated and the
specific heat becomes smoother. Moreover, different finite values ($> \log2/N$) of the
ground state entropy are found, signaling an exponential degeneracy of the
ground state.

\section{Conclusions}

In this paper we have developed a formalism capable of dealing efficiently
with Ising models defined on group lattices of non trivial genus $g$. In
particular we have applied the method to the Ising model on the Klein group $
L(2,7)$ having $g=3$ and $N=168$. We found that the computation of the
partition function ${\cal Z}$ is greatly simplified by use of
symmetries of an extended group, both internal and external to the group, 
which reduce the number
of and provide explicit formulas and topological interpretation for the sign
and weight of Pfaffians in the expansion of ${\cal Z}$. We plan to apply
this method to other lattices where $N$ is large and $g$ is comparable to $N$.

\section{Acknowledgments.}

Credit for ideas in this paper should be shared with Mario Rasetti to whom
we are indebted for many illuminating discussions and suggestions.

\appendix{\bf Appendix A}

Fermionic irreps of $H_0$ :

\begin{eqnarray*}
U^{(1)} &=&{\rm diag}(-1,K,-K^2,-K^4),\,U^{(2)}=\left( U^{(1)}\right)
^{*}\,,\,\, \\
U^{(3)} &=&{\rm diag}(K,-K^2,-K^4,-K^6,K^5,K^3)\, , \, \, U^{(4)}=\left(
U^{(3)}\right)^{*} \,,\,\, \\
U^{(5)}&=&{\rm diag} (K,-K^2,-K^4,-1,-K^6,K^5,K^3,-1)\,\, .
\end{eqnarray*}

\begin{equation}
V^{(1)}=\frac{2i}{\sqrt{7}}\left( 
\begin{array}{cccc}
1/2 & \frac 1{\sqrt{2}} & \frac 1{\sqrt{2}} & \frac 1{\sqrt{2}} \\ 
\frac 1{\sqrt{2}} & c_1 & c_3 & c_2 \\ 
\frac 1{\sqrt{2}} & c_3 & c_2 & c_1 \\ 
\frac 1{\sqrt{2}} & c_2 & c_1 & c_3
\end{array}
\right) \,\,\,, \\
\nonumber
\end{equation}
with $c_k=\cos \left( \frac{2k\pi }7\right) ,k=1,2,3$. $V^{(2)}=\left(
V^{(1)}\right) ^{*}$ .

\[
V^{(3)}=\frac{2i}7\left( 
\begin{array}{cc}
{\bf A} & {\bf B} \\ 
{\bf B} & -{\bf A}
\end{array}
\right) \,\,\,,
\]
where 
\begin{eqnarray}
{\bf A} &=&\left( 
\begin{array}{ccc}
s_2-\sqrt{2}s_3 & s_1-\sqrt{2}s_2 & s_3-\sqrt{2}s_1 \\ 
s_1-\sqrt{2}s_2 & s_3-\sqrt{2}s_1 & s_2-\sqrt{2}s_3 \\ 
s_3-\sqrt{2}s_1 & s_2-\sqrt{2}s_3 & s_1-\sqrt{2}s_2
\end{array}
\right) \;\;, \nonumber \\ 
{\bf B} &=&r\left( 
\begin{array}{ccc}
s_3+(\sqrt{2}-1)s_1 & s_2+(\sqrt{2}-1)s_3 & s_1+\sqrt{2}-1)s_2 \\ 
s_2+(\sqrt{2}-1)s_3 & s_1+\sqrt{2}-1)s_2 & s_3+(\sqrt{2}-1)s_1 \\ 
s_1+\sqrt{2}-1)s_2 & s_3+(\sqrt{2}-1)s_1 & s_2+(\sqrt{2}-1)s_3
\end{array}
\right) \,\,,
\nonumber 
\end{eqnarray}
with $s_1=\sin \left( \frac \pi 7\right) ,s_2=-\sin \left( \frac{2\pi }7
\right) ,s_3=-\sin \left( \frac{4\pi }7\right) $ and $r=\sqrt{2+\sqrt{2}}$ .

Finally $V^{(4)}=\left( V^{(3)}\right) ^{*}\,$and

\[
V^{(5)}=\frac i7\left( 
\begin{array}{cc}
{\bf C} & {\bf D} \\ 
{\bf D} & -{\bf C}
\end{array}
\right) \,\,\,,
\]
where
\[
{\bf C}=\left( 
\begin{array}{cccc}
s_1+s_2 & s_1+s_3 & s_2+s_3 & \sqrt{\frac 32}(s_2-s_1) \\ 
s_1+s_3 & s_2+s_3 & s_1+s_2 & \sqrt{\frac 32}(s_1-s_3) \\ 
s_2+s_3 & s_1+s_2 & s_1+s_3 & \sqrt{\frac 32}(s_3-s_2) \\ 
\sqrt{\frac 32}(s_2-s_1) & \sqrt{\frac 32}(s_1-s_3) & \sqrt{\frac 32}
(s_3-s_1) & \frac 12(s_1-2s_2+s_3)
\end{array}
\right) \;\;\;\;,
\]
and

\[
{\bf D}=2\sqrt{3}\left( 
\begin{array}{cccc}
s_2 & s_1 & s_3 & \frac 1{\sqrt{2}}(s_1-s_2^{}) \\ 
s_1 & s_3 & s_2 & \frac 1{\sqrt{2}}(s_3-s_1^{}) \\ 
s_3 & s_2 & s_1 & \frac 1{\sqrt{2}}(s_2-s_3^{}) \\ 
\frac 1{\sqrt{2}}(s_1-s_2^{}) & \frac 1{\sqrt{2}}(s_3-s_1^{}) & \frac 1{
\sqrt{2}}(s_2-s_3^{}) & \frac 12(s_1-s_3)
\end{array}
\right) \,\,\,\,\,.
\]

All these irreps are changed into equivalent and conjugate expressions by
replacing $K$ with $-K^2.$

\appendix{\bf Appendix B}

Putting $\lambda (x)=x_O$ for the representations $\lambda $ belonging to
the orbit $O$ we have

Case $B$:

\[
v_B=\left( 
\begin{array}{cc}
i & 0 \\ 
0 & -i
\end{array}
\right) \,\,\,\,,\,\,\,\, t_B=\left( 
\begin{array}{cc}
\frac{(1+i\,\sqrt{2})}2 & -\frac 12 \\ 
\frac 12 & \frac{(1-i\,\sqrt{2})}2
\end{array}
\right) \,\,\,\,\,\,\,,\,\,\,\,Z_{1,B}=\,{\bf I}_{2\,}\,;
\]

Case $C$: $\,\,\,\,\,\,\,v_C=v_B^{-1}\,\,\,,\,\,\,\,\,t_C=\,t_B^{-1}\,\,,\,
\,\,\,Z_{1,C}=\,Z_{1,B}\,$;

Case $D$: 
\begin{eqnarray*}
v_D^{(1)} &=&\frac 1{\sqrt{2}}\left( 
\begin{array}{cccccc}
0 & 0 & -1 & 1 & 0 & 0 \\ 
0 & 0 & -1 & -1 & 0 & 0 \\ 
1 & 1 & 0 & 0 & 0 & 0 \\ 
-1 & 1 & 0 & 0 & 0 & 0 \\ 
0 & 0 & 0 & 0 & -i & -i \\ 
0 & 0 & 0 & 0 & -i & i
\end{array}
\right) \,\,\,\,,\,\,\,\,\,\,t_D^{(1)}=\frac 1{\sqrt{2}}\left( 
\begin{array}{cccccc}
0 & 0 & \sqrt{2} & 0 & 0 & 0 \\ 
0 & 0 & 0 & \sqrt{2} & 0 & 0 \\ 
0 & 0 & 0 & 0 & -i & -i \\ 
0 & 0 & 0 & 0 & -i & i \\ 
-i & -i & 0 & 0 & 0 & 0 \\ 
-i & i & 0 & 0 & 0 & 0
\end{array}
\right) \,,\,\,\, \\
Z_{1,D}^{(1)} &=&{\rm diag}(1,1,1,1,-1,-1),\, \\
&&\,\, \\
&& \\
v_D^{(2)} &=&\left( v_D^{(1)}\right) ^{*}\,\,\,,\,\,\,t_D^{(2)}=\left(
t_D^{(1)}\right) ^{*}\,\,\,,\,\,Z_{1,D}^{(2)}\,=\,\,Z_{1,D}^{(1)}\text{ ;}
\end{eqnarray*}

Case $E$\thinspace :

\begin{eqnarray*}
v_E^{(1)} &=&\left( 
\begin{array}{cccc}
0 & 0 & -i & 0 \\ 
0 & -i & 0 & 0 \\ 
-i & 0 & 0 & 0 \\ 
0 & 0 & 0 & -i
\end{array}
\right) \,\,\,,\,\,\,t_E^{(1)}=\left( 
\begin{array}{cccc}
-1 & 0 & 0 & 0 \\ 
0 & 0 & -i & 0 \\ 
0 & 0 & 0 & 1 \\ 
0 & -i & 0 & 0
\end{array}
\right) \,\,\,,\,\, \\
Z_{1,E}^{(1)} &=&{\rm diag}(1,-1,1,-1)\,\,\,\,\,\,,\,v_E^{(2)}=\left(
v_E^{(1)}\right) ^{*}\,,\,\,\,\,\,\,t_E^{(2)}=\left( t_E^{(1)}\right)
^{*}\,\,\,\,,\,\,\,\,Z_{1,E}^{(2)}=-\,Z_{1,E}^{(1)}\,\,, \\
&& \\
&& \\
\,\,\,\,\,\,v_E^{(3)} &=&\left( 
\begin{array}{cccccccc}
0 & 0 & 0 & 0 & 1 & 0 & 0 & 0 \\ 
0 & 0 & 0 & 0 & 0 & 1 & 0 & 0 \\ 
0 & 0 & i & 0 & 0 & 0 & 0 & 0 \\ 
0 & 0 & 0 & -i & 0 & 0 & 0 & 0 \\ 
-1 & 0 & 0 & 0 & 0 & 0 & 0 & 0 \\ 
0 & -1 & 0 & 0 & 0 & 0 & 0 & 0 \\ 
0 & 0 & 0 & 0 & 0 & 0 & i & 0 \\ 
0 & 0 & 0 & 0 & 0 & 0 & 0 & -i
\end{array}
\right) \,\,,\,\,t_E^{(3)}=\left( 
\begin{array}{cccccccc}
\frac 12 & \frac{-\sqrt{3}}2 & 0 & 0 & 0 & 0 & 0 & 0 \\ 
\frac{\sqrt{3}}2 & \frac 12 & 0 & 0 & 0 & 0 & 0 & 0 \\ 
0 & 0 & 0 & 0 & 1 & 0 & 0 & 0 \\ 
0 & 0 & 0 & 0 & 0 & 1 & 0 & 0 \\ 
0 & 0 & 0 & 0 & 0 & 0 & \frac{-1}2 & \frac{-\sqrt{3}}2 \\ 
0 & 0 & 0 & 0 & 0 & 0 & \frac{-\sqrt{3}}2 & \frac 12 \\ 
0 & 0 & \frac 12 & \frac{\sqrt{3}}2 & 0 & 0 & 0 & 0 \\ 
0 & 0 & \frac{\sqrt{3}}2 & \frac{-1}2 & 0 & 0 & 0 & 0
\end{array}
\right) \,\,, \\
Z_{1,E}^{(3)} &=&{\rm diag}(-1,-1,1,1,-1,-1,1,1)\,\,\,.
\end{eqnarray*}

\vfill
\eject

\begin{figure}
\caption{Intersection between  $\ell$ and $\ell'$  and 
schematic representation of $\zeta, \zeta'$ and $\zeta+\zeta'$ (or $\ell, \ell'$ and $\ell''$).}
\label{fig1}
\end{figure}

\begin{figure}
\caption{Specific heat versus temperature for $J_U=J_V=1$.}
\label{fig2}
\end{figure}

\vfill


\begin{references}
\bibitem{note1} Not all determinants can be computed symbolically
but all of them can be evaluated numerically by very fast algorithms.
\end{references}
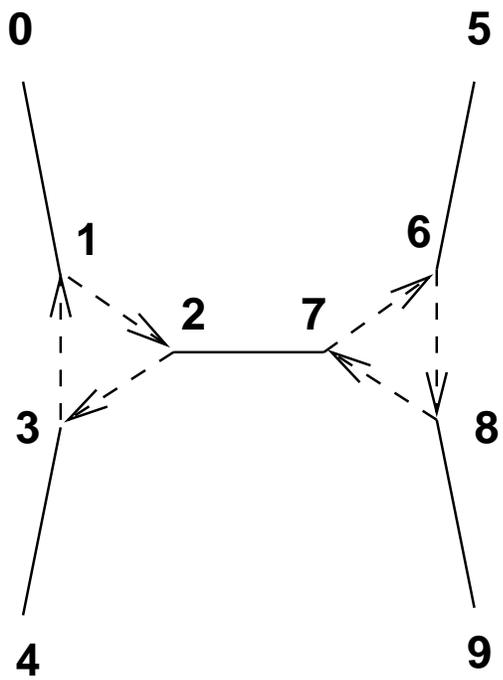
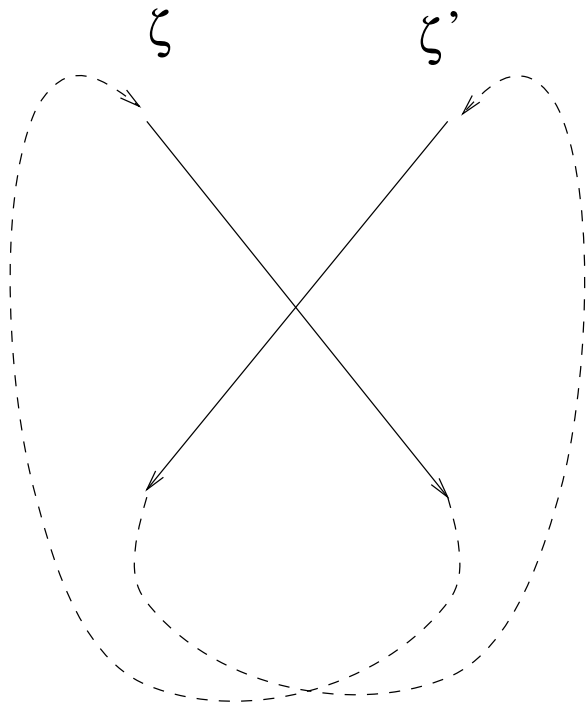
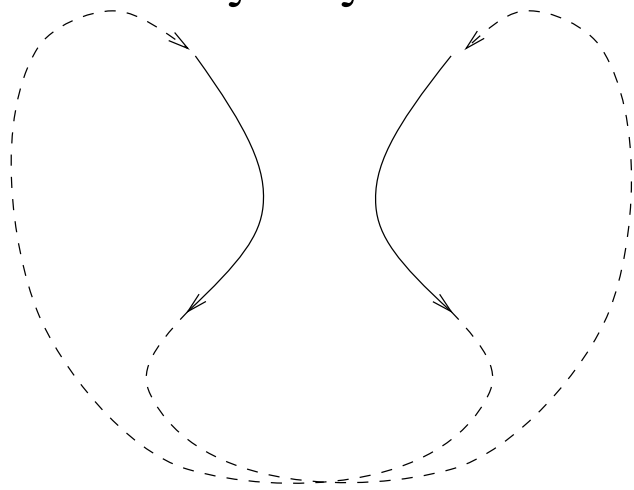
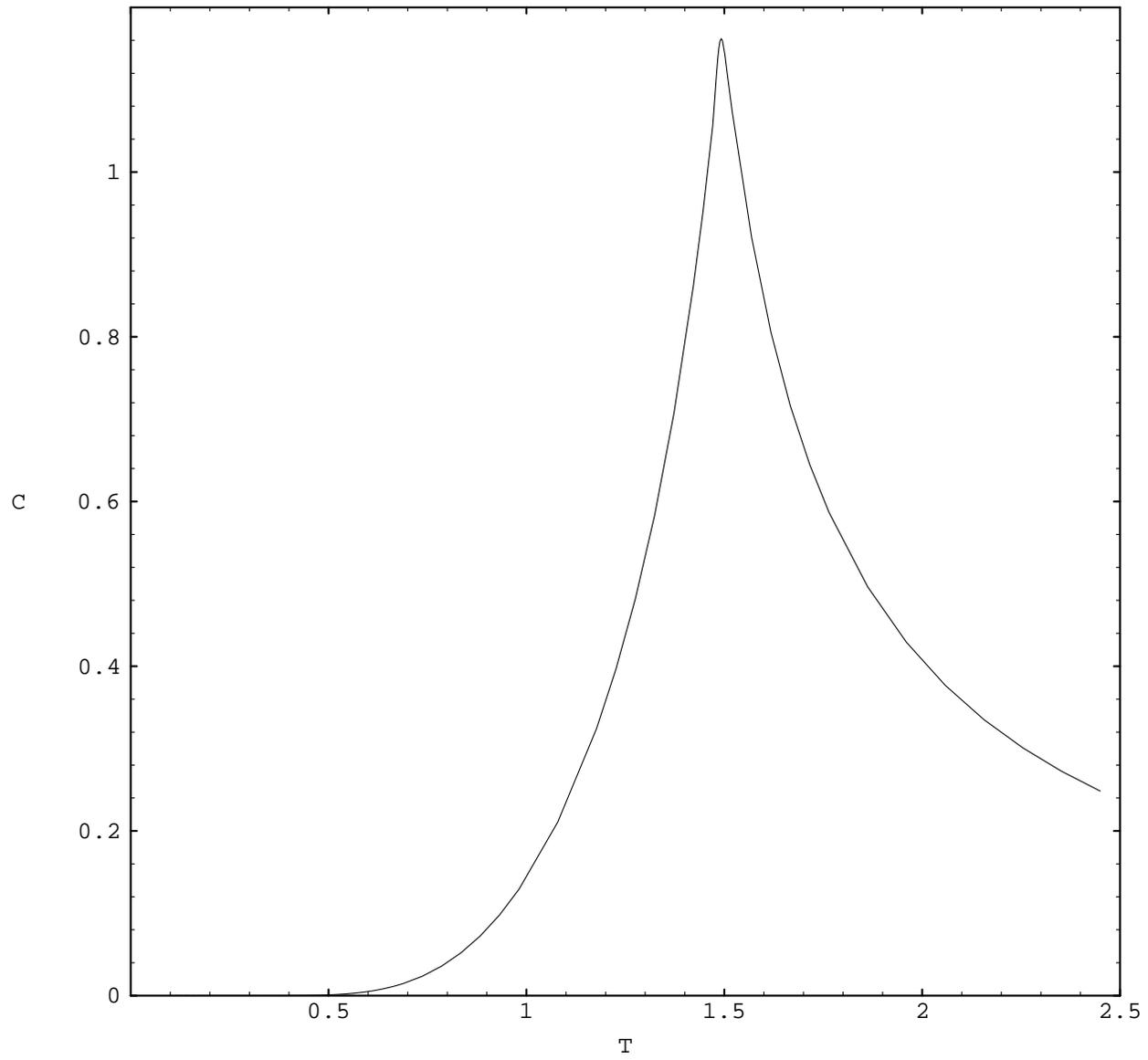
\end{document}